\documentclass{article}
\usepackage{spconf,amsmath,graphicx, booktabs}
\usepackage{multirow}
\usepackage{color}

\title{A Study on Bias and Fairness In Deep Speaker Recognition}
\name{Amirhossein Hajavi\thanks{The authors would like to thank IMRSV Data Labs for their support of this work. The authors would also like to acknowledge the Natural Sciences and Engineering Research Council of Canada (NSERC) for supporting this research (grant number: CRDPJ 533919-18).}, Ali Etemad}
\address{\textit{Dept. ECE and Ingenuity Labs Research Institute,}\\Queen's University, Canada \\
\{a.hajavi, ali.etemad\}@queensu.ca}
\begin{document}
\ninept
\maketitle
\begin{abstract}
With the ubiquity of smart devices that use speaker recognition (SR) systems as a means of authenticating individuals and personalizing their services, fairness of SR systems has becomes an important point of focus. In this paper we study the notion of fairness in recent SR systems based on 3 popular and relevant definitions, namely \textit{Statistical Parity}, \textit{Equalized Odds}, and \textit{Equal Opportunity}. We examine 5 popular neural architectures and 5 commonly used loss functions in training SR systems, while evaluating their fairness against gender and nationality groups. Our detailed experiments shed light on this concept and demonstrate that more sophisticated encoder architectures better align with the definitions of fairness. Additionally, we find that the choice of loss functions can significantly impact the bias of SR models.
\end{abstract}
\begin{keywords}
Deep speaker recognition, bias, fairness.
\end{keywords}
\section{Introduction}
\label{sec:intro}
Speaker recognition (SR) systems are being used in a variety of different smart devices for identifying or authenticating users. Their uses include granting access to individuals \cite{duraibi2020voice} who intend to use the products/services provided by the smart devices or customize the provided services by personalizing the experience towards each user \cite{gourav2021personalization}. 
Recently, deep neural networks (DNN) have become the predominant mechanism used in SR systems \cite{hajavi2019deep, Nagrani19voxceleb, chung2020in, hajavi2021siamese, hajavi2022fine,  jung2022pushing}. This is mainly due to factors such as improved performance in comparison to traditional SR techniques as a result of state-of-the-art neural architectures and loss functions used in training \cite{bai2021speaker}. 

In recent years, it has been well-reported that deep learning models are susceptible to biases, which leads to `unfair' decision making \cite{mehrabi2021survey}. An unfair SR system may perform well in favor of a particular group of people while performing poorly for others. On the other hand, a `fair' SR system \textit{avoids any favoritism towards any particular group based on the characteristics of that group} \cite{mehrabi2021survey}. In case of smart devices, an unfair SR system could result in the device being only usable by \textit{some} people while rendering it useless for other groups of people. In the context of fairness, the group of people that may receive disadvantages from the decisions made by a biased system are referred to as the `protected' group \cite{verma2018fairness} and the other groups are referred to as `unprotected'.


In this paper we evaluate recent SR systems in terms of \textit{fairness} and report the results of our evaluation using 3 widely used definitions of fairness.
For this study, we select 5 of the most popular neural encoders which are frequently used in SR systems.
Moreover, to consider the impact of different loss functions towards fairness, we train each model with 5 different common loss functions.
The protected groups in this study are defined as `gender' and `nationality' of speakers. The VoxCeleb dataset \cite{Nagrani19voxceleb} is used in this study given that it is the most widely used dataset in training SR systems and the wide range of diversity among its speakers.

In summary we make the following contributions:
    (\textbf{1}) We evaluate fairness of the current widely used architectures in SR and cross-examine them with different loss functions used in training. 
    (\textbf{2}) Our study provides a new and comprehensive perspective on fairness in SR systems by incorporating several popular encoder architectures, comparing different methods of training, and evaluating the impact of different factors of bias (both gender and nationality). 
    (\textbf{3}) We report the results of our experiments in the form of a comparative analysis that shows the impact of using each combination of architecture/loss function on fairness of SR systems. 

{The remainder the paper is organized as follows. First, we briefly review existing works in the area. Next, we provide a detailed explanation of definitions of fairness used in our study, followed by our study design, including a description of the architectures of the networks used. Further, the loss functions used in the training of SR systems are described. In the following section, we provide the details of the performed experiments, including the dataset and training details. Finally, we report the results of our experiments and provide a comparative discussion on these results. Lastly, we conclude our study by presenting a summary of the work.}

\section{Related Work}
To the best of our knowledge, a comprehensive study of fairness in SR systems has been studied in two prior works in the literature \cite{Hutiri2022spkr_fairness, toussaint2021sveva}. 
Both works evaluate 2 ResNet-based models, namely ResNet34V2 and ResNet34L in their studies. Although they provide a thorough analysis on the performance of these 2 DNNs with respect to the protected and unprotected groups, their work does not consider all the factors that are involved in the training of the DNNs, for instance loss functions. Furthermore, these works only take 1 definition of fairness into consideration, despite the availability of a variety of different definitions for fairness \cite{verma2018fairness}.


Other studies such \cite{fenu2020improv} and \cite{shen2022improv} aim to \textit{improve} the fairness of SR systems. The study done in \cite{fenu2020improv} studies the effect of modifying the training set of SR systems to balance the representations of the protected and unprotected groups. The work done in \cite{shen2022improv} proposes a novel fusion network by training multiple DNNs. Each network is trained using data from 1 group. A different model is trained to receive scores from these networks and calculate a fusion score which is then used for making the final decision. While these studies help the fairness of the SR systems, 
they do not analyse the fairness of SR models with regards to different architectures and training procedures. {In this paper, we aim to address and extend these shortcomings to include the effect of different architectures and training methods on fairness of SR systems.}

\section{Methods}
\label{sec:method}

In this section, we describe the different components of our study design including definitions of fairness, SR neural architectures, and training loss functions.

\subsection{Fairness Definitions}

The notion of fairness has been widely studied, and a number of different formulations have been proposed to capture its essence \cite{verma2018fairness}.
In this study, we select 3 popular definitions that we believe to be relevant to our study:
\textit{Statistical parity}, \textit{Equalized odds}, and \textit{Equal opportunity}. All these 3 formulations depend on statistical measures, which can be defined by the output of SR systems being either `positive' (the test and reference utterances are spoken by the same person) or `negative'. Following we provide the detailed formulations of these 3 definitions.



\noindent{\textbf{\textit{Statistical parity.}} 
The system meets the requirements for statistical parity when the probability of a positive decision made by the system for both protected and unprotected groups are equal. This definition can be formulated using: 
    \begin{equation}
        P(d=1|A=1) - P(d=1|A=0) = 0,
    \end{equation}
where $P(d=1|A=1)$ is the probability of positive decision $d=1$ for the protected group $A=1$ and $P(d=1|A=0)$ is the same probability for the unprotected group.
}

\noindent{\textbf{\textit{Equalized odds.}} 
The system satisfies the definition of equalized odds when the system has an equal probability of making correct positive decisions and equal probability of making incorrect positive decisions, for both the protected and unprotected groups. This can be formulated as: 
\begin{equation}
        P(d=1|A=1, Y) - P(d=1|A=0, Y) = 0, 
\end{equation}
where the ground truth labels are shown by $Y$ and can take values of 0 or 1.
}

\noindent{\textbf{\textit{Equal opportunity.}} 
The system satisfies the definition of equal opportunity if the probability of incorrect negative decisions made by the system is equal for both the protected and unprotected groups. This can be formulated as: 
\begin{equation}
\small
        P(d=0|A=1, Y=1) - P(d=0|A=0, Y=1) = 0.
\end{equation}
}

{In the formulations above, probability values higher than 0 indicate the presence of bias in the learned model. These values can be directly related to the amount of bias in SR system with higher values indicating more bias in the systems.}

\subsection{Speaker Recognition Systems}
\begin{figure}[t]
  \centering
  \includegraphics[width=1\linewidth]{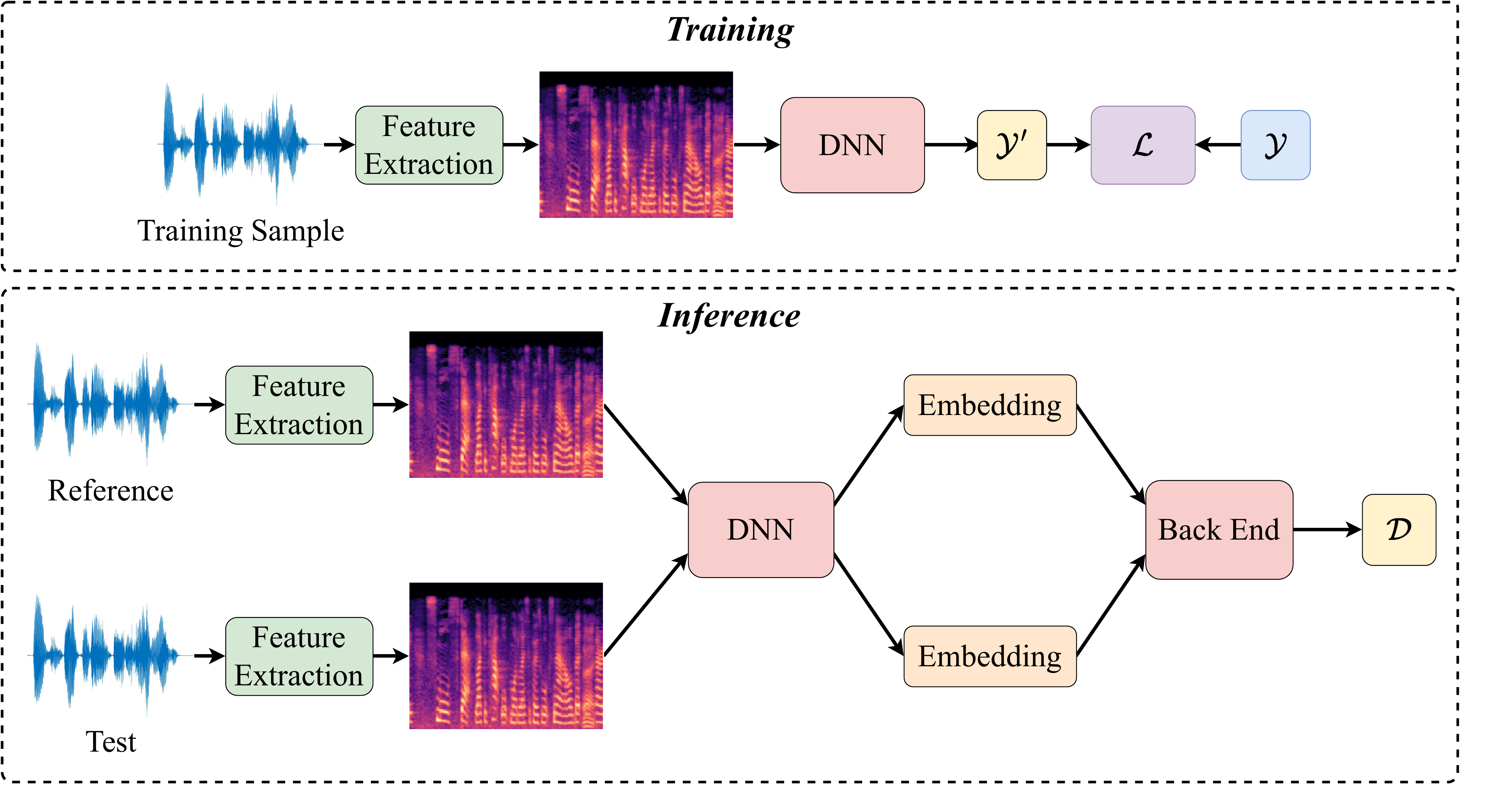}
  \caption{The general training and inference pipelines of a speaker recognition system.}
  \label{fig:SR}
\end{figure}

Figure \ref{fig:SR} (Top) portrays the training pipeline of a general SR system. The first step in the pipeline is extraction of frequency features from raw audio signals. The extracted features, namely, spectrograms are then passed onto a DNN. The prediction made by the DNN, denoted in the figure by $\mathcal{Y'}$, is then compared to the ground-truth labels, denoted in the figure by $\mathcal{Y}$, using the loss function $\mathcal{L}$. The main components of this pipeline that have attracted the most interest in the recent literature are the DNN and $\mathcal{L}$ \cite{chung2020in}. During the inference stage, shown in Figure \ref{fig:SR} (Bottom), the DNN component generates embeddings from 2 utterances, namely, reference and test. The Backend component then makes a decision $\mathcal{D}$ by comparing the embeddings and calculating a cosine distance of the embeddings. In order to evaluate the performance of SR systems, Equal Error Rate (EER) is used which is calculated by measuring the error rate after selecting a threshold on the calculated cosine distance. The threshold is selected at a point where the number of incorrect negative decisions and incorrect positive decisions are equal to each other.

\begin{table*}[!t]
\resizebox{1\textwidth}{!}{%
\begin{tabular}{l|l|l|l||l|l||l|l|l||l|l||l|l|l||l|l}
\multirow{2}{*}{DNN} & \multicolumn{5}{c||}{Statistical parity}   & \multicolumn{5}{c||}{Equalized odds} & \multicolumn{5}{c}{Equal opportunity} \\ \cline{2-16}
                     & $\mathcal{L}_S$ & $\mathcal{L}_{AMS}$ & $\mathcal{L}_{AAMS}$ & $\mathcal{L}_{T}$ & $\mathcal{L}_{P}$  &       $\mathcal{L}_S$ & $\mathcal{L}_{AMS}$ & $\mathcal{L}_{AAMS}$ & $\mathcal{L}_{T}$ & $\mathcal{L}_{P}$  &      $\mathcal{L}_S$ & $\mathcal{L}_{AMS}$ & $\mathcal{L}_{AAMS}$ & $\mathcal{L}_{T}$ & $\mathcal{L}_{P}$         \\ \hline\hline
VGG-M-40             & 0.151       &  0.143      &  0.138     &  0.159      &  0.141         &  0.098     &  0.091     &  0.088     &  0.102    &  0.087     &  0.081     &  0.078      &  0.073      &  0.086     &  0.075       \\
ResNet34L            & 0.147       &  0.139      &  0.135     &  0.152      &  0.140         &  0.085     &  0.078     &  0.074     &  0.092    &  0.081     &  0.072     &  0.070      &  0.063      &  0.077     &  0.068      \\
ResNet34V2           & 0.126       &  0.121      &  0.119     &  0.132      &  0.121         &  0.079     &  0.075     &  0.072     &  0.085    &  0.075     &  0.065     &  0.062      &  0.057      &  0.068     &  0.061       \\
SEResNet34L          & 0.136       &  0.131      &  0.127     &  0.141      &  0.130         &  0.082     &  0.078     &  0.075     &  0.091    &  0.075     &  0.074     &  0.068      &  0.061      &  0.081     &  0.065       \\
SEResNet34V2         & 0.131       &  0.128      &  0.123     &  0.134      &  0.127         &  0.079     &  0.076     &  0.074     &  0.088    &  0.075     &  0.070     &  0.061      &  0.058      &  0.072     &  0.061       \\
\end{tabular}
}
\caption{Results of evaluation of SR systems on definitions of fairness by taking gender as the basis for defining the protected group}
\label{tab:results}
\end{table*}

\subsubsection{DNN architectures}
We study 5 different architectures that are used widely in the recent literature. These 5 models are based on VGG, ResNet, and SEResNet family of networks. Following, we present a brief description of each of the 5 DNNs. 

\noindent\textbf{VGG-M-40:} The VGG-based \cite{chatfield2014return} model studied in this work is proposed in \cite{nagrani2017voxceleb}. This model consists of 5 convolution layers, 3 maxpooling layers, an average-pooling layer, and 3 fully connected (FC) layers. Additionally, the model uses temporal average pooling (TAP) to aggregate the embeddings generated by the last FC layer across time. 

\noindent\textbf{ResNet34L:} The next architecture studied in this paper is based on the thin-ResNet model used in \cite{cai2018exploring, chung2020delving}, and is proposed in \cite{chung2020in}. It uses 34 convolution layers incorporated into residual blocks. Each residual block consists of 3 convolution layers, each coupled with a batch-normalization layer. The input of each residual block is connected to its output through a shortcut connection. The difference between ResNet34L and the original thin-ResNet is the smaller input and larger strides in the early layers of the model. Also the model uses self-attentive pooling (SAP) for aggregation of the embeddings across time. 

\noindent\textbf{ResNet34V2:} The second ResNet-based model reviewed in this study is proposed in \cite{kwon2021ins}. Similar to ResNet34L, the architecture details of this model is also based on thin-ResNet model used in \cite{cai2018exploring, chung2020delving}. However, unlike ResNet34L the input shape of the DNN has not changed compared to thin-ResNet and the model uses smaller strides as well. Additionally, the model utilizes attentive statistical pooling (ASP) for aggregation of embeddings.

\noindent\textbf{SEResNet34L:} The SEResNet-Based model studied in this paper is proposed in \cite{hajavi2022fine} and is a modified version of ResNet34L \cite{chung2020in}. In this model the residual blocks of the ResNet34L have been replaced by Squeeze-and-Excitation (SE) \cite{hu2018squeeze} blocks. The SE blocks, in addition to the shortcut connections, utilize an average pooling mechanism to obtain channel information from the input of the block and generate an excitation score for each channel. The generated score is then multiplied over the respective channel.

\noindent\textbf{SEResNet34V2:} Similar to SEResNet34L, this model is a modified version of ResNet34V2 \cite{kwon2021ins}. The residual blocks in ResNet34V2 model are replaced by the SE blocks.

\subsubsection{Loss functions}
We study 5 most commonly used loss functions in the recent literature \cite{chung2020in} for training SR systems. These functions can be divided into two categories of classification and metric learning \cite{chung2020in}. 
{The general difference between classification and metric learning loss functions lies in their consideration of intra-speaker relations while training the models. This difference raises an interest towards their effect on fairness of SR models trained by each of them.} {Following we describe these categories in more detail.}

\noindent\textbf{Classification.}
{In cases where the SR system is trained using a classification loss, the last layer of the DNN is an FC layer with $M$ nodes where each node represents one speaker from a predefined set of speakers. Given a batch of $N$ utterances, the DNN then generates latent embeddings $X$ prior to the last layer. }
{In this category of loss functions we select the popular and widely used Softmax, AM-Softmax, and AAM-Softmax functions.}

\noindent\textbf{Metric Learning.}
{In this category, SR systems aim to learn a distance or similarity function between pairs of speakers in the dataset. The goal of these models is therefore to optimize the metric so that it maps utterances of the same speaker closer together and dissimilar data points further apart in high-dimensional space. To evaluate the notion of bias in this category of SR systems, we consider two popular losses, namely Triplet loss and Prototypical loss.}

\noindent\textit{Triplet loss:} {In triplet loss \cite{schroff2015facenet}, the DNN is given 3 utterances. Among the given utterances, 2 utterances (referred to as anchor and positive) are spoken by the same speaker and 1 utterance (referred to as negative) is spoken by a different speaker. Triplet loss then minimizes the distance between the embeddings generated by the DNN for the anchor $x_a$ and positive utterances $x_p$, while maximizing the distance for $x_a$ and the negative utterance $x_n$ using:}
\begin{equation}
    \mathcal{L}_{T} = \max(0, \lVert x_a - x_p \lVert_{2}^{2} - \lVert x_a - x_n \lVert_{2}^{2})
\end{equation}

\noindent\textit{Prototypical loss:} {In prototypical loss \cite{snell2017porotypical}, at each step of training, a subset of $N$ speakers are selected. A support set is then created by selecting $m$ utterances for each speaker of the subset. Among the utterances selected for speaker $i$, $m-1$ utterances are used to create a prototype $c_i$ from embeddings generated by the network $x_{i,j}$ by: }
\begin{equation}
    c_{i} = \frac{1}{m-1}\sum\limits_{j=1}^{m-1} x_{i,j}.
\end{equation}
{The squared Euclidean distance between the remaining utterance $x_{i,m}$ and the prototype $c_k$ is then calculated using: }
\begin{equation}
    S_{i,k} = \lVert x_{i,m} - c_k \lVert_{2}^{2}.
\end{equation}
{The prototypical loss $\mathcal{L}_P$ is finally calculated by: }
\begin{equation}
    \mathcal{L}_{P} = -\frac{1}{N}\sum\limits_{i=1}^{N}\log\frac{e^{S_{i,i}}}{\sum_{k=1}^{N}e^{S_{i,k}}}.
\end{equation}

\section{Experiments and Results}
In this section, we first give a detailed description of the dataset used for evaluating the SR systems in terms of fairness and discuss the protected and unprotected groups defined for our experiments. We then describe the implementation details and training protocols. Finally, we present the results of the experiments and report our observations.

\subsection{Dataset}
We use the Voxceleb2 \cite{Nagrani19voxceleb} for training our models. This dataset is the most widely used dataset for training SR systems. At the time of this study, the dataset contained utterances from more than 7000 speakers. The utterances are collected from open-source media and are recorded in uncontrolled environments. For evaluation purposes, we use VoxCeleb1 \cite{Nagrani19voxceleb}. This dataset contains utterances from 1251 speakers from 11 different nationalities. 

\begin{figure*}[!t]
    \centering
    \includegraphics[width=1.\textwidth]{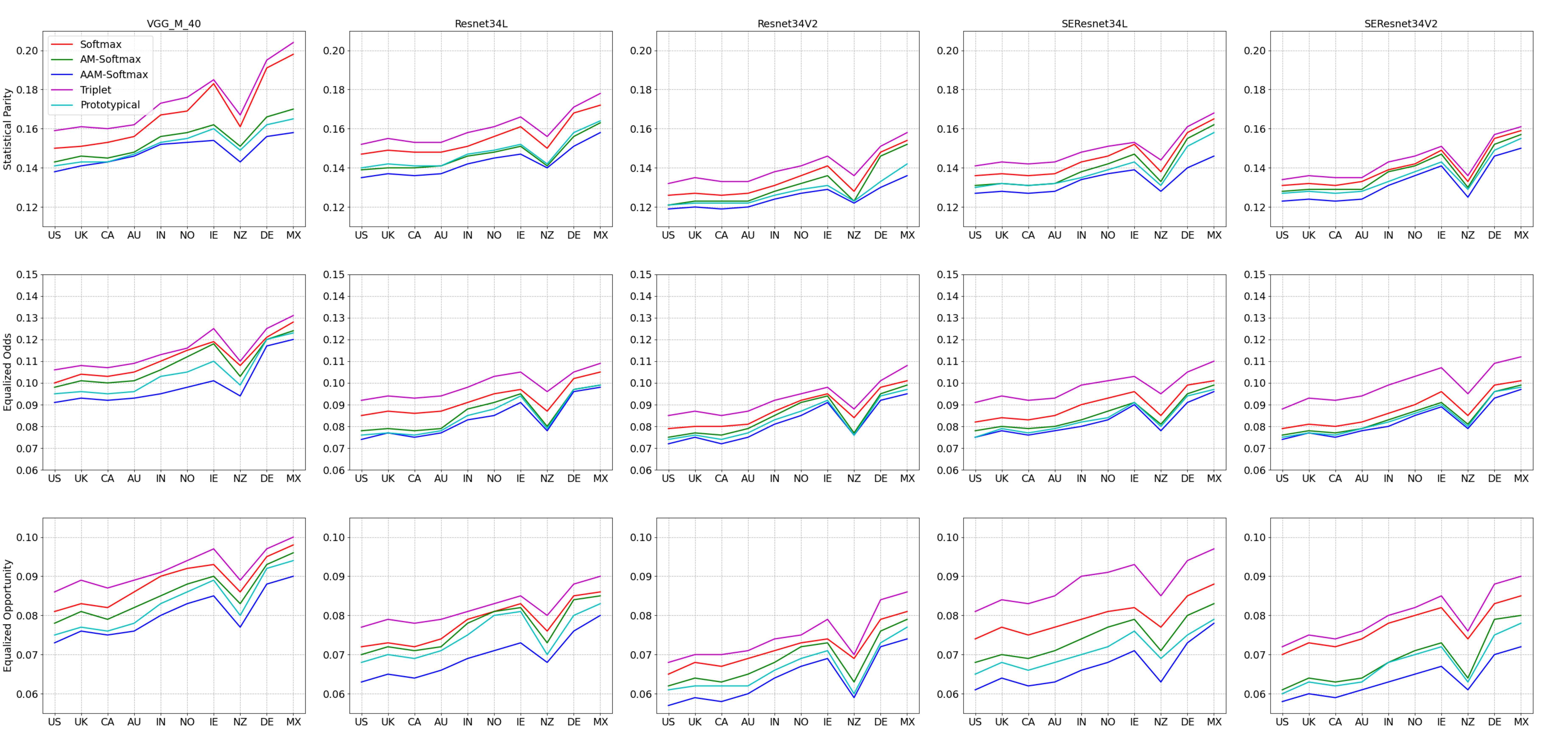}
    \caption{Results for evaluation of the SR systems on definitions of fairness by taking nationality as the basis for defining the protected groups}
    \label{fig:results_nationality}
\end{figure*}

In this study, we perform 2 sets of experiments by taking either gender or nationality as the basis of defining the protected and unprotected groups. In the first set of experiments we consider the female group of speakers as the protected group and perform fairness evaluation based on the 3 definitions of fairness. 
The measurements for the male group can be calculated as the complimentary value of the female group.
For the second set of experiments we consider the nationality of the speaker as the basis of defining the protected and unprotected groups. We take each nationality as a protected group and accumulate all the other nationalities together as the unprotected group.

\subsection{Training}
We use Adam optimizer for training the SR models. We use a step decay for scheduling the learning rate during the training of the models, and set the initial learning rate to $10^{-3}$. We train the models on a single Nvidia Titan RTX (24 GB vRAM) GPU, and set the batch-size to 64.

\subsection{Results}
Table \ref{tab:results}, shows the result of our experiments for evaluation of fairness in SR systems when taking gender as the basis to define the protected and unprotected groups. We measure the fairness metrics for systems using the same thresholds that are used when calculating EER in order to evaluate the fairness of the models with the same conditions that are present during inference. The results show that the ResNet34V2 architecture better satisfies the definitions of fairness followed by SEResNet34V2, SEResNet34L, ResNet34L, and finally VGG-M-40. By comparing ResNet34V2 and ResNet34L {in terms of number of parameters (see Table \ref{table:hyp})} we can observe that \textbf{\textit{using larger and more complex models {seems to reduce} the amount of bias in SR systems}}. This is further confirmed by comparing SEResNet34V2 and SEResNet34L or by comparing any of the larger models with VGG-M-40. 
This may be due the fact that contrary to the belief that larger models are more prune to over-fitting and thus being more biased, the popular models used in SR when trained with Voxceleb2 do not reach the over-fitting state during the training. We can also observe that using SE modules instead of normal residual blocks has no consistent impact on bias.

The result of the experiments also show that \textbf{\textit{the choice of loss functions can significant impact fairness in SR systems}}. We see the least amount of bias in systems that are trained by AAM-Softmax loss followed by the prototypical loss function. The highest amount of bias is present in the systems trained by Triplet-loss and Softmax loss functions. Comparing the fairness of all the models trained using AAM-Softmax with the models trained using AM-Softmax and regular Softmax suggests that by using angular margin the bias of the system can be reduced. Furthermore by comparing the prototypical loss function with triplet-loss function, it can be observed that using multiple prototypes in training of the SR systems instead of only 1 positive and 1 negative sample, results in a considerably less bias in the system.

\begin{table}[!t]
\centering
\begin{tabular}{l l} 
\hline
Method & Num. of parameters\\
\hline
  ResNet34L &  1.4M \\ 
  ResNet34V2 & 2.0M\\
  SEResNet34L & 1.4M \\
  SEResNet34V2 & 2.0M \\
\hline
\end{tabular}
\caption{{Comparison between the number of parameters on ResNet and SE-ResNet-based models.}}
\label{table:hyp}
\end{table}

Next, we repeat the experiments by defining the protected groups based on nationality of the speakers. 
We take each nationality to be a protected group while all the other nationalities are combined together as the unprotected group. Figure \ref{fig:results_nationality}, presents the results of this experiment. We sort the nationality of the speakers based on the number of speakers from each nationality. Comparing the fairness of the systems based on nationality suggests the presence of more bias while working on nationalities with fewer number of speakers, as expected. However, the bias seems to be commonly lower for 5 countries of US, UK, CA, AU, and NZ despite the number of speakers for NZ being lower. This may be because of the common characteristics (such as language and word pronunciations) of the speakers of these nationalities. 

Lastly, similar to the experiment with gender based groups, we observe that larger and more complex models such as ResNet34V2 contain the least amount of bias followed by SEResNet34V2, SEResNet34L, ResNet34L, and VGG-M-40. We can also see that the models trained using AAM-Softmax loss show the least amount of bias followed by prototypical loss, AM-Softmax loss, Softmax loss, and Triplet-loss.

\section{Conclusion}
Considering the ubiquity of smart devices that use SR systems, fairness of these systems is of major importance. This paper presents a comprehensive study on fairness in SR systems using 3 definitions of fairness, namely Statistical parity, Equalized odds, and Equal opportunity. We study 5 of the most popular DNN architectures and 5 different loss functions that are most commonly used in SR systems. We choose gender and nationality for defining the protected and unprotected groups. Our experiments show that larger models, such as ResNet34V2 and SEResNet34V2, trained with AAM-Softmax and prototypical loss functions better comply with the definitions of fairness. 

\bibliographystyle{IEEEbib}
\bibliography{ 
               refs, 
               refs_self
             }

\begin{thebibliography}{10}

\bibitem{duraibi2020voice}
Salahaldeen Duraibi,
\newblock ``Voice biometric identity authentication model for iot devices,''
\newblock {\em International Journal of Security, Privacy and Trust Management
  (IJSPTM)}, vol. 9, 2020.

\bibitem{gourav2021personalization}
Aditya Gourav, Linda Liu, Ankur Gandhe, Yile Gu, Guitang Lan, Xiangyang Huang,
  Shashank Kalmane, Gautam Tiwari, Denis Filimonov, Ariya Rastrow, et~al.,
\newblock ``Personalization strategies for end-to-end speech recognition
  systems,''
\newblock {\em IEEE International Conference on Acoustics, Speech and Signal
  Processing (ICASSP)}, pp. 7348--7352, 2021.

\bibitem{hajavi2019deep}
Amirhossein Hajavi and Ali Etemad,
\newblock ``A deep neural network for short-segment speaker recognition,''
\newblock {\em {INTERSPEECH}}, pp. 2878--2882, 2019.

\bibitem{Nagrani19voxceleb}
Arsha Nagrani, Joon~Son Chung, Weidi Xie, and Andrew Zisserman,
\newblock ``Voxceleb: Large-scale speaker verification in the wild,''
\newblock {\em Computer Science and Language}, 2019.

\bibitem{chung2020in}
Joon~Son Chung, Jaesung Huh, Seongkyu Mun, Minjae Lee, Hee~Soo Heo, Soyeon
  Choe, Chiheon Ham, Sunghwan Jung, Bong-Jin Lee, and Icksang Han,
\newblock ``In defence of metric learning for speaker recognition,''
\newblock {\em INTERSPEECH}, 2020.

\bibitem{hajavi2021siamese}
Amirhossein Hajavi and Ali Etemad,
\newblock ``Siamese capsule network for end-to-end speaker recognition in the
  wild,''
\newblock {\em IEEE International Conference on Acoustics, Speech and Signal
  Processing (ICASSP)}, pp. 7203--7207, 2021.

\bibitem{hajavi2022fine}
Amirhossein Hajavi and Ali Etemad,
\newblock ``Fine-grained early frequency attention for deep speaker
  recognition,''
\newblock {\em arXiv preprint arXiv:2207.10006}, 2022.

\bibitem{jung2022pushing}
Jee-weon Jung, You~Jin Kim, Hee-Soo Heo, Bong-Jin Lee, Youngki Kwon, and
  Joon~Son Chung,
\newblock ``Pushing the limits of raw waveform speaker recognition,''
\newblock {\em INTERSPEECH}, 2022.

\bibitem{bai2021speaker}
Zhongxin Bai and Xiao-Lei Zhang,
\newblock ``Speaker recognition based on deep learning: An overview,''
\newblock {\em Neural Networks}, vol. 140, pp. 65--99, 2021.

\bibitem{mehrabi2021survey}
Ninareh Mehrabi, Fred Morstatter, Nripsuta Saxena, Kristina Lerman, and Aram
  Galstyan,
\newblock ``A survey on bias and fairness in machine learning,''
\newblock {\em ACM Computing Surveys (CSUR)}, vol. 54, no. 6, pp. 1--35, 2021.

\bibitem{verma2018fairness}
Sahil Verma and Julia Rubin,
\newblock ``Fairness definitions explained,''
\newblock {\em IEEE/ACM international workshop on software fairness
  (fairware)}, pp. 1--7, 2018.

\bibitem{Hutiri2022spkr_fairness}
Wiebke~Toussaint Hutiri and Aaron~Yi Ding,
\newblock ``Bias in automated speaker recognition,''
\newblock {\em ACM Conference on Fairness, Accountability, and Transparency},
  p. 230–247, 2022.

\bibitem{toussaint2021sveva}
Wiebke Toussaint and Aaron~Yi Ding,
\newblock ``Sveva fair: A framework for evaluating fairness in speaker
  verification,''
\newblock {\em arXiv preprint arXiv:2107.12049}, 2021.

\bibitem{fenu2020improv}
Gianni Fenu, Giacomo Medda, Mirko Marras, and Giacomo Meloni,
\newblock ``Improving fairness in speaker recognition,''
\newblock {\em European Symposium on Software Engineering}, p. 129–136, 2020.

\bibitem{shen2022improv}
Hua Shen, Yuguang Yang, Guoli Sun, Ryan Langman, Eunjung Han, Jasha Droppo, and
  Andreas Stolcke,
\newblock ``Improving fairness in speaker verification via group-adapted fusion
  network,''
\newblock {\em ICASSP 2022 - 2022 IEEE International Conference on Acoustics,
  Speech and Signal Processing (ICASSP)}, pp. 7077--7081, 2022.

\bibitem{chatfield2014return}
Ken Chatfield, Karen Simonyan, Andrea Vedaldi, and Andrew Zisserman,
\newblock ``Return of the devil in the details: Delving deep into convolutional
  nets,''
\newblock {\em arXiv preprint arXiv:1405.3531}, 2014.

\bibitem{nagrani2017voxceleb}
Arsha Nagrani, Joon~Son Chung, and Andrew Zisserman,
\newblock ``Voxceleb: A large-scale speaker identification dataset,''
\newblock {\em Proc. Interspeech 2017}, pp. 2616--2620, 2017.

\bibitem{cai2018exploring}
Weicheng Cai, Jinkun Chen, and Ming Li,
\newblock ``Exploring the encoding layer and loss function in end-to-end
  speaker and language recognition system,''
\newblock {\em arXiv preprint arXiv:1804.05160}, 2018.

\bibitem{chung2020delving}
Joon~Son Chung, Jaesung Huh, and Seongkyu Mun,
\newblock ``Delving into voxceleb: Environment invariant speaker recognition,''
\newblock {\em Odyssey: The Speaker and Language Recognition Workshop}, pp.
  349--356, 2020.

\bibitem{kwon2021ins}
Yoohwan Kwon, Hee-Soo Heo, Bong-Jin Lee, and Joon~Son Chung,
\newblock ``The ins and outs of speaker recognition: lessons from voxsrc
  2020,''
\newblock {\em IEEE International Conference on Acoustics, Speech and Signal
  Processing (ICASSP)}, pp. 5809--5813, 2021.

\bibitem{hu2018squeeze}
Jie Hu, Li~Shen, and Gang Sun,
\newblock ``Squeeze-and-excitation networks,''
\newblock {\em {IEEE conference on computer vision and pattern
  recognition(CVPR)}}, pp. 7132--7141, 2018.

\bibitem{schroff2015facenet}
Florian Schroff, Dmitry Kalenichenko, and James Philbin,
\newblock ``Facenet: A unified embedding for face recognition and clustering,''
\newblock {\em IEEE/CVF conference on computer vision and pattern recognition
  (CVPR)}, pp. 815--823, 2015.

\bibitem{snell2017porotypical}
Jake Snell, Kevin Swersky, and Richard Zemel,
\newblock ``Prototypical networks for few-shot learning,''
\newblock {\em Advances in Neural Information Processing Systems (NeurIPS)},
  vol. 30, 2017.

\end{thebibliography}

\end{document}